# Adaptive nonreciprocal wave attenuation in linear piezoelectric metastructures shunted with one-way electrical transmission lines

Yisheng Zheng, Junxian Zhang, Yegao Qu*, Guang Meng

State Key Laboratory of Mechanical System and Vibration, Shanghai Jiao Tong University, 200240, Shanghai, China

*Corresponding author
 Email: yszheng@sjtu.edu.cn (Y. Zheng), quyegao@sjtu.edu.cn (Y. Qu).

**Abstract:** In contrary to elastic media, it is easy to attain one-way coupling feature among electrical elements, which enables unidirectional transmission of electrical wave. In this paper, we explore and exploit the interaction of a piezoelectric beam with a one-way electrical transmission line to facilitate nonreciprocal transmission of elastic wave in the linear fashion. Theoretical dispersion analysis and numerical simulations are performed to reveal transmission behaviors of elastic wave in the presented piezoelectric metastructure. It is uncovered that, when the one-way electrical coupling feature is introduced, local-resonance bandgaps are maintained in opposite directions of the piezoelectric metastructure. However, the bandgap attenuation capability in one direction is increased compared to conventional local-resonance metastructures, in which no electrical coupling exists between cells, while that in the other direction is decreased. Therefore, nonreciprocal transmission of elastic wave emerges in this system due to the distinct bandgap attenuation capability. It is also revealed that this sort of nonreciprocal wave attenuation capability is adaptive with the one-way electrical coupling coefficient and the resistance of electrical cells. An experimental set-up of the piezoelectric metastructure is built and experimental results verify the nonreciprocity property and its adaptiveness. Overall, the presented piezoelectric metastructure provides a linear and concise approach to realize adaptive nonreciprocal transmission of elastic wave.

## 1. Introduction

Nonreciprocity of elastic/acoustic wave is an emerging research topic, which could play important roles in the areas such as underwater acoustic communication, vibration energy control and ultrasonic imaging [1,2]. In regular elastic media, the transmission of elastic/acoustic wave obeys the reciprocity theorem [3]. There have been many efforts trying to break this limitation and achieve unidirectional transmission of elastic/acoustic wave, among which, the scheme utilizing nonlinear properties of elastic media is the first one proposed [4]. The nonlinear approach may be the most popular one investigated for realizing nonreciprocity to date. All sorts of nonlinear nonreciprocal systems have been explored, for instance: Liang et al. [5] presented an acoustic rectifier which couples a superlattice with a nonlinear medium, and experimentally observed significant rectifying effect within the predicted frequency bands; Boechler et al. [6] utilized the bifurcation mechanism to design a nonlinear granular crystal with rectification ratios greater than $10^4$; Lepri and Casati [7] studied a layered nonlinear, nonmirror-symmetric model for nonreciprocal wave transmission; Popa and Cummer [8] designed a highly nonlinear nonreciprocal device composed of two Helmholtz resonators and a piezoelectric membrane with nonlinear circuits shunting; Wu et al. [9] and Zheng et al. [10] reported nonreciprocal mechanical metastructures and piezoelectric metastructures designed based on the nonlinear wave transmission phenomenon called "supratransmission", respectively. Though the nonlinear approach offers many opportunities in designing nonreciprocal systems, it usually brings the drawback of distorting frequencies of wave energy, which however is an important characteristic needed to be preserved during wave propagation, especially in applications where the wave is regarded as signal rather than just energy. Besides, the nonlinear approach has restrictions on excitation amplitudes that could induce nonreciprocity. Therefore, it is significant to investigate linear elastic/acoustic systems to achieve nonreciprocity. Fleury et al. [11] performed a pioneer work on the linear acoustic circulator with nonreciprocity. The spatial-temporal modulation is also known as an effective linear approach to achieve nonreciprocity. Trainiti and Ruzzene [12] studied analytically the longitudinal and transverse elastic wave propagation behaviors in a periodic beam with spatial-temporal modulated properties, and found that directional bandgaps existed in such systems. Nassar et al. [13] characterize analytically how wave-like modulation leads to dispersion curves transformation in spatial-temporal phononic crystals, and revealed the relationship between reciprocity breaking and Willis coupling effects. Recently, the spatial-temporal modulation technique has also been utilized to realize nonreciprocity of Rayleigh wave [14,15]. Though several experimental works have been reported [16–18], the physical implementation of spatial-temporal nonreciprocal systems is relatively complicated, since special control techniques are required to modulate physical parameters such as modulus and density in both the space and time domain. In addition, the active control strategy provides another path to facilitate nonreciprocity. Several important works in this area were reported recently such as that, Baz [19] proposed a eigen-structure assignment control strategy to design a nonreciprocal acoustic duct, Sasmal et al. [20] studied a non-local active metamaterials with broadband acoustic nonreciprocity, and Chen et al. [21] presented an active mechanical Willis meta-layer with highly asymmetric wave transmission ratios. To date, the research on active nonreciprocal elastic/acoustic metamaterials is still at the preliminary stage. Generally, the existing mechanisms and designs that could achieve linear nonreciprocity are very limited. It remains challenging to design concise and linear nonreciprocal elastic/acoustic systems that could be easily implemented physically in engineering applications.

To put nonreciprocity into applications, since the practical environment and requirements may be constantly changing, the adaptiveness is an important feature worth addressing. To tackle this challenge, piezoelectric systems have superiority due to its flexibility of performance tuning, which can be realized by conveniently altering shunting circuits or control strategies [22]. Since first proposed by Thorp et al. [23], piezoelectric metamaterials have proved great effectiveness and adaptiveness in manipulating elastic wave. Sugino et al. [24] reported a digitally programmable piezoelectric metamaterial that enables simultaneous tuning over the central frequency, attenuation and bandwidth of local-resonance bandgaps. Xu and Tang [25] designed a tunable prism featured with continuous beam steering of elastic wave. Darabi et al. [26] experimentally implemented a reconfigurable electroacoustic topological insulator. By dynamically tuning shunting circuits to modulate effective mechanical properties of piezoelectric metamaterials such as density and modulus, new opportunities of wave manipulation can also be attained. For instance, Trainiti et al. [27] presented a piezoelectric waveguide using switching circuits to implement time-modulated stiffness for selective wave filtering, whilst Marconi et al. [18] and Sugino et al. [28] studied nonreciprocal bandgaps facilitated in space-time modulated piezoelectrical beams [21,28]. In addition to the aforementioned piezoelectric metamaterials that are all shunted with local circuits, piezoelectric metamaterials shunted with electrical transmission lines, namely interconnected circuits, have also been studied for elastic wave and vibration control. Lossouarn et al. [29] proposed to utilize the coupling of a mechanical lattice with an electrical network to achieve multimodal vibration suppression. Flores Parra et al. [30] studied dispersion crossing effects in a piezoelectric metamaterial with *L-C* high pass and bandpass electrical networks. Overall, piezoelectric metamaterials have played important roles in realizations of exceptional and adaptive wave propagation properties. It could serve as an appropriate physical platform to attain adaptive elastic/acoustic nonreciprocity.

In this research, we present a novel piezoelectric metastructure shunted with a one-way electrical transmission line to realize unidirectional transmission of elastic wave in the linear and adaptive fashion. Distinct from existing piezoelectric metamaterials, which with either local shunting circuits or two-way electrical transmission lines, the presented electrical-mechanical system features a one-way electrical transmission line, which consists of periodic electrical cells that are neighborly coupled in the one-way fashion. By investigating elastic wave transmission behaviors of this system theoretically and experimentally, we found that local-resonance bandgaps still exist in such piezoelectric metamaterials even though the shunting circuits are not locally interacted with the mechanical beam. We also found that the wave attenuation capability in the bandgap is direction-dependent, which thus facilitates nonreciprocal transmission of elastic wave. This nonreciprocity mechanism is different from the spatial-temporal modulation technique, where the bandgap frequency is direction-dependent. It is also revealed that nonreciprocal transmission ratios of the presented system are adaptive with the one-way electrical coupling coefficient and the resistance of electrical cells. Since the presented system attains nonreciprocity in local-resonance bandgaps, it has the subwavelength control capability of elastic wave. The physical implementation of the presented linear and adaptive nonreciprocal system is relatively simple since it does not need special modulation and control techniques. This makes it a promising approach to attain linear nonreciprocity and could have great application potentials.

The paper is organized as follows. In Section 2, the physical model of the presented nonreciprocal piezoelectrical metastructure is illustrated. Dispersion analysis of the metastructure cell is conducted in Section 3, followed by numerical simulations of the finite piezoelectric metastructure in Section 4. Additionally, the experimental set-up of the nonreciprocal system is described and the experimental results are explained in Section 5. Lastly, some concluding remarks are summarized in Section 6.

## 2. Description of the electrical-mechanical system

In this section, the physical model of the piezoelectric metastructure is described. Since the one-way electrical transmission line is a crucial part of the system, it would be introduced and discussed firstly, followed by illustrations of the whole electrical-mechanical system.

### *2.1 One-way electrical transmission line*

The common linear physical systems usually support wave propagation reciprocally. To attain nonreciprocity, special design techniques are required. Actually, we can readily achieve nonreciprocity if one-way physical couplings are attainable. Figure 1. (a) shows the schematic of a one-way coupling physical system, where neighboring two media are coupled in the one-way fashion, viz. the response of medium $j$ ($j$=1, 2…$N$-1) can affect that of medium $j$+1 but not the reverse. In such a system, the physical wave propagates from medium 1 to medium $N$ unidirectionally. However, based on the Newton's third law, which states that the reaction force equals to the action force, it is hard to attain the one-way coupling feature in conventional elastic media. Thus, it is difficult to achieve nonreciprocal elastic wave transmission.

On the contrary, it is much easier to attain the one-way coupling feature in electrical systems. A sort of one-way coupling electrical element is presented in the dashed box of Fig. 1(b), which is an inverting amplifier and couples Circuit 1 with Circuit 2 in the one-way fashion. Herein, the resistance $R_a$ is chosen to be much larger than the impedance of Circuit 1 and therefore the current across $R_a$ can be neglected. Under this condition, this one-way coupling electrical element can be seen as a voltage-controlled voltage source (VCVS). To ensure that the current across $R_a$ can be neglected, we can also connect a buffer circuit in series with the inverting amplifier at node 1. In Fig. 1(b), the voltage response $V_2$ depends on $V_1$ in the relation

$$V_2 = G \cdot V_1 \qquad (1)$$

where $G=-R_b/R_a$. The electrical response $V_1$ is solely determined by Circuit 1 and has no dependence on Circuit 2, which is the cause of one-way coupling feature. Actually, there are many designs that can realize the one-way electrical coupling feature. Without loss of generality, we use the presented one, due to its simplicity, to demonstrate the nonreciprocity property and mechanism of the piezoelectric metastructure in this paper. It is worth noting that the electrical diode is also a nonreciprocal device but only used to attain nonreciprocity for DC signals. Utilizing one-way coupling electrical elements, we could conveniently establish one-way electrical transmission lines allowing nonreciprocal transmission of electrical wave.

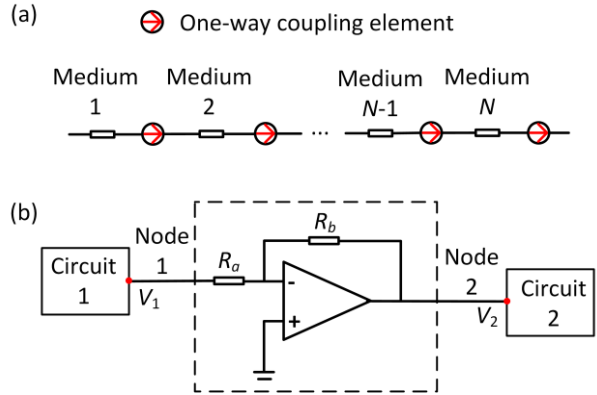

**Fig. 1.** (a) Schematic of physical systems with unidirectional wave transmission, neighboring media are coupled in the one-way fashion. (b) Diagram of one-way coupling electrical element (in the dashed box), which couples Circuit 1 and Circuit 2 in the one-way fashion

*2.2 Physical model of piezoelectrical metastructure*

The electrical-mechanical cell equipped with the one-way electrical coupling element is illustrated in Fig. 2 (a). It is composed of two segments $a$ and $b$, where the substrate layer of segment $a$ is covered by two PZT patches, which are polarized in $y$ direction and connected in the parallel configuration. The PZT patches are shunted with an electric cell, of which the one-way coupling electrical element is shown in Fig. 1(b). For the electrical cell, according to Eq. (1), $U_m$ and $U_{n+1}$ is governed by the formulation

$$U_{n+1} = G \cdot U_m \qquad (2)$$

If the one-way electrical coupling coefficient $G=0$, the voltage $U_n=U_{n+1}=0$. Then the electrical cell becomes a local shunting circuit with inductor $(L_1+L_2)$ in series with $(R_1+R_2)$. In this scenario, the electrical-mechanical system would degenerate into a conventional local-resonance piezoelectric metamaterial. And the existence of $L_1$ and $L_2$ could help formulate local-resonance bandgaps in this electrical-mechanical system.

The geometrical and physical parameters of the metastructure cell are illustrated in the following. As shown in Fig. 2(a), the length of cells is $x_{n+1}-x_n=l_c$, where $x_n$ and $x_{n+1}$ are the positions of left end and right end of $n$th cell, respectively. The width and height of the substrate layer are $b_b$ and $h_b$, respectively. The length of PZT patches is $x_m-x_n=l_p$ with $x_m$ being the position of right end of the PZT patch. The width and height of PZT patches are respectively $b_p$ and $h_p$. The density and Young's modulus of the Aluminum substrate layer are respectively $E_b$ and $\rho_b$, while that of PZT patches are respectively $E_p$ and $\rho_p$. The piezoelectric strain constant is $d_{31}$. The dielectric constant is $\varepsilon_{33}^T$. The capacitance of PZT patches at constant strain is $C_p$ for each cell.

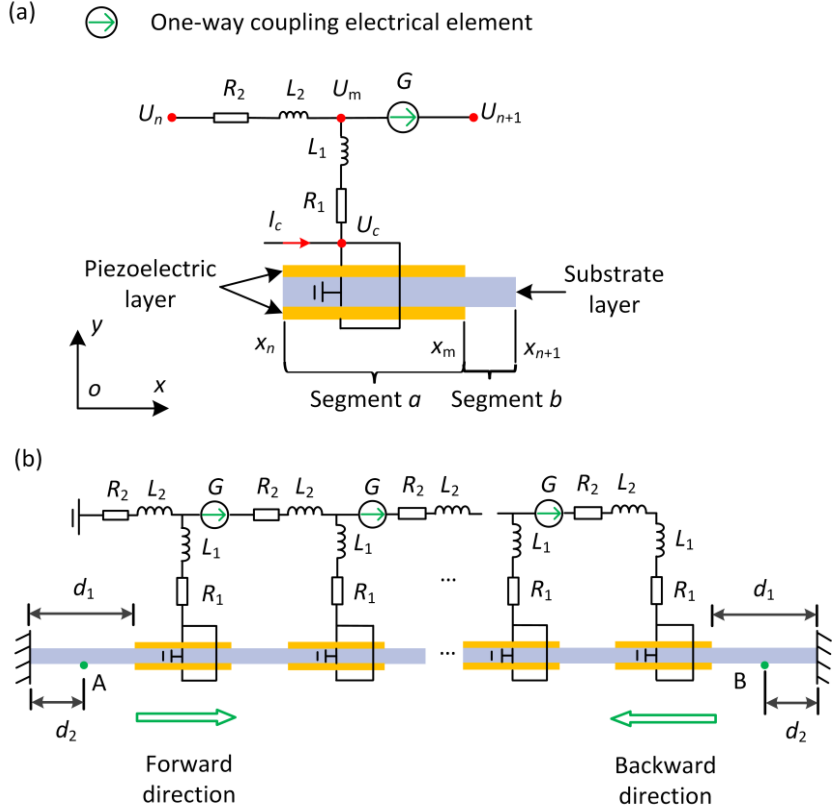

**Fig. 2.** (a) Unit-cell of piezoelectric metastructure; the diagram of one-way coupling electrical element is given as Fig. 1(b); $I_c$ is the current generated by PZT patches. (b) Finite nonreciprocal piezoelectric metastructures shunted with one-way electrical transmission lines: for studying forward wave transmission, the force $F$ is applied at point A and the acceleration $a_s$ of point B is measured; for studying backward wave transmission, the force $F$ is applied at point B and the acceleration $a_s$ of point A is measured.

The schematic of the piezoelectric metastructure with finite number of cells is presented in Fig. 2(b), where the metastructure is placed on the central part of a beam with the fixed-fixed end boundary condition. The total length of the beam is $d$. The piezoelectric metastructure has $N_c$ cells, which is located at the position $d_1$ from the ends of the beam. The two points A and B, located at the position $d_2$ from ends of the beam, are designated for actuating and sensing. In the electrical transmission line, neighboring cells are coupled in the one-way fashion by using one-way coupling electrical elements presented as Fig. 1(b). The electrical system is grounded on the left end while the right end is open. It is readily recognized that elastic wave transmission in the piezoelectric beam alone is reciprocal while electric wave propagates nonreciprocally in the electrical transmission line alone. However, due to interaction between the piezoelectric beam and the electrical transmission line, wave transmission properties in the coupled electrical-mechanical system are more complex and deserves further investigations.

To reveal nonreciprocity of the presented piezoelectric metastructure in Fig. 2(b), we would investigate wave transmission properties in both directions. When the forward wave transmission is considered, the force $F$ is applied at point A and the acceleration $a_s$ of point B is measured. In contrast, when the backward wave transmission is studied, the force $F$ is applied at point B and the acceleration $a_s$ of point A is measured. The acceleration response can be used to evaluate the transmission capability of elastic wave.

## 3. Dispersion relation of electrical-mechanical cell

The dispersion diagrams are usually employed to analyze wave propagation properties of metamaterials. In this section, the analytical study of the electrical-mechanical cell would be conducted by calculating its dispersion relations. There are two methods to obtain the dispersion diagrams [31], either calculating wavenumbers with given frequencies (known as the direct method) or calculating eigenfrequencies with given wavenumbers (known as the inverse method). Using the direct method, the calculated imaginary part of wavenumber is related to the wave attenuation constants, which is favor for characterizing nonreciprocal wave attenuation capability of the proposed system and therefore would be utilized in this paper.

*3.1 Analytical modelling of cell*

To calculate dispersion relations of the proposed electrical-mechanical system, the transfer matrix method [32,33] would be employed. Considering the Euler beam assumption, the dynamic state variables of the beam is given as $\mathbf{Y}=[w\ w_x\ M\ Q]^\mathrm{T}$. $w$ and $w_x$ are the transverse displacement and its derivative with respective to the position $x$, respectively; $M$ and $Q$ are the bending torque and shear force, respectively. For the beam segment $a$ with PZT layers, as shown in Fig. 2(a), the dynamic equations in frequency domain are presented as

$$\begin{cases} \dfrac{\mathrm{d}w}{\mathrm{d}x}=w_x,\ D_a\dfrac{\mathrm{d}w_x}{\mathrm{d}x}=M+\lambda U_c \\ \dfrac{\mathrm{d}M}{\mathrm{d}x}=-Q,\ \dfrac{\mathrm{d}Q}{\mathrm{d}x}=-\omega^2 m_a w \end{cases} \quad (3)$$

where $D_a$ and $m_a$ are the bending rigidity and linear mass density of segment $a$, respectively, given as

$$\begin{cases} D_a=E_b b_b h_b^3/12+E_p b_p[(h_b+2h_p)^3-h_b^3]/12 \\ m_a=\rho_b b_b h_b+2\rho_p b_p h_p \end{cases} \quad (4)$$

$U_c$ is the voltage on PZT patches. The electrical-mechanical coupling coefficient $\lambda$ is given as [34]

$$\lambda = E_p d_{31} b_p \cdot (h_p + h_b) \tag{5}$$

Eq. (3) can be rearranged into the matrix form as

$$\mathbf{Y}' = \mathbf{C}_a \cdot \mathbf{Y} + \mathbf{J}_a \cdot U_c \tag{6}$$

with

$$\mathbf{C}_a = \begin{bmatrix} 0 & 1 & 0 & 0 \\ 0 & 0 & 1/D_a & 0 \\ 0 & 0 & 0 & -1 \\ -\omega^2 m_a & 0 & 0 & 0 \end{bmatrix} \mathbf{Y}, \quad \mathbf{J}_a = \begin{bmatrix} 0 \\ \lambda/D_a \\ 0 \\ 0 \end{bmatrix} \tag{7}$$

For a first-order differential equation in the matrix form presented as

$$\mathbf{Y}'(x) = \mathbf{C} \cdot \mathbf{Y}(x) + \mathbf{Y}_0 \tag{8}$$

where $\mathbf{Y}$ is a vector variable, $\mathbf{C}$ is a constant matrix and $\mathbf{Y}_0$ is a constant vector, its general solution is written as

$$\mathbf{Y} = \gamma e^{\mathbf{C}x} - \mathbf{C}^{-1} \cdot \mathbf{Y}_0 \tag{9}$$

where $\gamma$ is an arbitrary coefficient. Therefore, the solution of Eq. (6) could be obtained as

$$\mathbf{Y} = \gamma e^{\mathbf{C}_a x} - \mathbf{C}_a^{-1} \cdot \mathbf{J}_a \cdot U_c \tag{10}$$

from which, we can readily derive the relation between $\mathbf{Y}$ at position $x_m$ and that at position $x_n$, given as

$$\mathbf{Y}_m + \mathbf{C}_a^{-1} \cdot \mathbf{J}_a \cdot U_c = \mathbf{F} \cdot (\mathbf{Y}_n + \mathbf{C}_a^{-1} \cdot \mathbf{J}_a \cdot U_c) \tag{11}$$

where $\mathbf{F} = e^{\mathbf{C}_a \cdot l_p}$.

As for segment $b$, letting $U_c=0$ in Eq. (6) and replacing $\mathbf{C}_a$ with $\mathbf{C}_b$, we can obtain the similar formulation for segment $b$, namely

$$\mathbf{Y}' = \mathbf{C}_b \cdot \mathbf{Y} \tag{12}$$

$\mathbf{C}_b$ is achieved by simply replacing $D_a$ and $m_a$ in Eq. (7) with $D_b$ and $m_b$, respectively, which are presented as

$$\begin{cases} D_b = E_b b_b h_b^3 / 12 \\ m_b = \rho_b b_b h_b \end{cases} \tag{13}$$

Similar with Eq. (11), the relation between $\mathbf{Y}$ at position $x_{n+1}$ and that at position $x_m$ is then given as

$$\mathbf{Y}_{n+1} = \mathbf{H} \cdot \mathbf{Y}_m \tag{14}$$

where $\mathbf{H} = e^{\mathbf{C}_b \cdot (l_c - l_p)}$.

Substituting Eq. (14) into Eq. (11) to eliminate $\mathbf{Y}_m$, we can derive the relation between $\mathbf{Y}$ at position $x_n$ and that at position $x_{n+1}$, given as

$$\mathbf{Y}_{n+1} = \mathbf{H} \cdot \mathbf{F} \cdot \mathbf{Y}_n + \mathbf{B} \cdot U_c \tag{15}$$

where $\mathbf{B}$ is written as

$$\mathbf{B} = (\mathbf{H} \cdot \mathbf{F} - \mathbf{H}) \cdot \mathbf{C}_a^{-1} \cdot \mathbf{J}_a \tag{16}$$

The matrix $\mathbf{B}$ characterizes the electrical-mechanical coupling effect.

For the electrical cell shown in Fig. 2(a), the following two dynamic equations can be formulated

$$U_c = (1 + Z_1/Z_2)/G \cdot U_{n+1} - Z_1/Z_2 \cdot U_n \tag{17}$$

$$(Z_1 + Z_2 + Z_c)/G \cdot U_{n+1} = (Z_1 + Z_c) \cdot U_n + Z_c Z_2 \cdot I_c \tag{18}$$

where $Z_1 = L_1 s + R_1$, $Z_2 = L_2 s + R_2$, $Z_c = 1/(C_p s)$. $s$ is the complex number and $s = j\omega$, with j being the imaginary unit and $\omega$ being the angular frequency. In Eq.(18), the current $I_c$ generated by PZT patches in each cell is [34]

$$I_c = -s\lambda \mathbf{J}_b \cdot (\mathbf{Y}_m - \mathbf{Y}_n) \tag{19}$$

where $\mathbf{J}_b = [0\ 1\ 0\ 0]$. $\mathbf{Y}_m$ is related to $\mathbf{Y}_{n+1}$ and can be derived from Eq. (14).

Substituting Eq. (17) into Eq. (15) for eliminating the variable $U_c$ leads to

$$\mathbf{Y}_{n+1} - (1 + Z_1/Z_2)/G \cdot \mathbf{B} \cdot U_{n+1} = \mathbf{H} \cdot \mathbf{F} \cdot \mathbf{Y}_n - Z_1/Z_2 \cdot \mathbf{B} \cdot U_n \tag{20}$$

By combining Eqs. (14)(18) and (19), it is derived that

$$s\lambda Z_c Z_2 \cdot \mathbf{J}_b \cdot \mathbf{H}^{-1} \mathbf{Y}_{n+1} + (Z_1 + Z_2 + Z_c)/G \cdot U_{n+1} = s\lambda Z_c Z_2 \cdot \mathbf{J}_b \cdot \mathbf{Y}_n + (Z_1 + Z_c) \cdot U_n \tag{21}$$

Eqs. (20) and (21) can be rewritten in the matrix form, which relates the general state variables of the electrical-mechanical cell at left end and that at right end, presented as

$$\mathbf{X}_{n+1} = \mathbf{T}_c \cdot \mathbf{X}_n \tag{22}$$

with $\mathbf{X}_n = [\mathbf{Y}_n\ U_n]^T$, $\mathbf{X}_{n+1} = [\mathbf{Y}_{n+1}\ U_{n+1}]^T$ and $\mathbf{T}_c = \mathbf{P}^{-1} \cdot \mathbf{Q}$. $\mathbf{T}_c$ is the unit-cell transfer matrix. The terms $\mathbf{P}$ and $\mathbf{Q}$ are presented as

$$\begin{cases} \mathbf{P} = \begin{bmatrix} \mathbf{I}_{4\times 4} & -(1 + Z_1/Z_2)/G \cdot \mathbf{B} \\ s\lambda Z_c Z_2 \cdot \mathbf{J}_b \cdot \mathbf{H}^{-1} & (Z_1 + Z_2 + Z_c)/G \end{bmatrix} \\ \mathbf{Q} = \begin{bmatrix} \mathbf{H} \cdot \mathbf{F} & -Z_1/Z_2 \cdot \mathbf{B} \\ s\lambda Z_c Z_2 \cdot \mathbf{J}_b & Z_1 + Z_c \end{bmatrix} \end{cases} \tag{23}$$

According to the Floquet-Bloch theorem, $\mathbf{X}_n$ can be written as $\mathbf{X}_n = \mathbf{A}_x \cdot e^{jkx_n + j\omega t}$ with $\mathbf{A}_x$ being the plane wave amplitude. Thus the relation between $\mathbf{X}_n$ and $\mathbf{X}_{n+1}$ could be written as

$$\mathbf{X}_{n+1} = e^{jk \cdot l_c} \mathbf{X}_n \tag{24}$$

where $k$ is the wavenumber. From Eqs. (22) and (24), it is achieved that

$$(\mathbf{T}_c - e^{jk \cdot l_c} \mathbf{I}) \cdot \mathbf{X}_n = 0 \tag{25}$$

where $\mathbf{I}$ is the unit matrix. To acquire non-zero solutions of $\mathbf{X}_n$ from Eq. (25), we have

$$\left| \mathbf{T}_c - e^{jk \cdot l_c} \mathbf{I} \right| = 0 \tag{26}$$

which represents the dispersion relation of the electrical-mechanical cell. For a given frequency $\omega$, the calculated $k$ using Eq. (26) could be expressed in the complex form. The imaginary part of wavenumber $k$ determines the wave attenuation constant $\mu$ of the metastructure based on the following expression

$$\mu=\text{Im}(k)\cdot l_c \qquad (27)$$

Therefore, the value Im($k$) can reflect the wave attenuation capability of the system.

*3.2 Dispersion analysis*

With Eq. (26), we can then calculate dispersion diagrams of the electrical-mechanical metastructure cell. The geometrical and physical parameters of the cell are given as Table 1. The electrical parameters are presented as follows: $L_1$=0.1 H, $L_2$=0.1 H, $R_1$=200 Ω and $R_2$=200 Ω.

Table 1. Geometrical and physical parameters of the piezoelectric metastructure cell

| Parameters | Values | Parameters | Values |
| --- | --- | --- | --- |
| Substrate-layer density $\rho_b$ | 2700 [Kg/m$^3$] | Piezo-layer density $\rho_p$ | 7780 [Kg/m$^3$] |
| Substrate-layer modulus $E_b$ | 69 [GPa] | Piezo-layer modulus $E_p$ | 60 [GPa] |
| Unit cell length $l_c$ | 17 [mm] | Piezo-layer length $l_p$ | 15 [mm] |
| Substrate-layer width $b_b$ | 15 [mm] | Piezo-layer width $b_p$ | 15 [mm] |
| Substrate-layer height $h_b$ | 2 [mm] | Piezo-layer height $h_p$ | 1 [mm] |
| Dielectric constant $e_{33}^T$ | 1.6×10$^{-8}$ [F/m] | Piezoelectric strain constant $d_{31}$ | -1.7×10$^{-10}$ [C/N] |
| PZT capacitance $C_p$ | 6.4 [nF] | / | / |

Before analyzing dispersion relations of the coupled electrical-mechanical system, we first calculate dispersion curves of the uncoupled system by assuming the piezoelectric strain constant $d_{31}$ to be 0, which can be used as the benchmark for analyzing dispersion relations of the coupled system. The calculated results are presented in Fig. 3, where the same type lines in Fig. 3(a) and (b) represent the same dispersion mode. The pink lines represent electrical wave modes, where two cases $G$=-0.001 and $G$=-0.1 are considered. These two cases have identical Re($k$) but different Im($k$). The cyan and red lines represent propagating elastic wave modes in opposite directions while the green and blue lines represent evanescent elastic wave modes in opposite directions. For the propagating elastic wave modes, the imaginary part of wavenumber Im($k$)=0. For the evanescent elastic wave modes, the real part of wavenumber Re($k$)=0. Since there is no electrical-mechanical coupling effect, no bandgap phenomenon is observed from the dispersion curves of elastic wave. As seen from the electrical wave modes, when the electrical one-way coupling coefficient is chosen as $G$=-0.001, Im ($k$) is very large, which indicates that electrical wave decays drastically in the electrical transmission line. When $G$ is increased to be -0.1, Re ($k$) remains unchanged, while Im($k$) is decreased and thus the electrical wave transmission capability is improved.

To get a comprehensive understanding of the electrical wave propagation behaviors, we are herein to determine the natural frequencies of the finite electrical transmission line shown in Fig. 2(b). For a *L-C* circuit, its natural frequency is known as $1/(2\pi\sqrt{L\cdot C})$. The natural frequencies of common electrical transmission lines [29,30] with the two-way coupling feature are distinct from that of local electrical cells. However, for the presented one-way electrical transmission line, its natural frequencies may have different characteristics and are worth further studying. Considering the electrical cell without coupling to the structure, Eq. (18) is reduced to

$$(Z_1+Z_c)\cdot U_n-(Z_1+Z_2+Z_c)/G\cdot U_{n+1}=0 \qquad (28)$$

Base on Eq. (28), the equation of the whole electrical transmission line is then presented as

$$\begin{cases} (Z_1 + Z_c) \cdot U_1 - (Z_1 + Z_2 + Z_c)/G \cdot U_2 = 0 \\ (Z_1 + Z_c) \cdot U_2 - (Z_1 + Z_2 + Z_c)/G \cdot U_3 = 0 \\ \vdots \\ (Z_1 + Z_c) \cdot U_{Nc} - (Z_1 + Z_2 + Z_c)/G \cdot U_{Nc+1} = 0 \end{cases} \quad (29)$$

To calculate the natural frequencies, we need to apply certain boundary conditions. As presented in Fig. 2(b), the left-end of the electrical transmission line is short while its right end is open, which leads to $U_1$=0. Therefore, we have the following formulation from Eq. (29)

$$\begin{bmatrix} a & & & & \\ b & a & & & \\ & b & a & & \\ & & b & \ddots & \\ & & & \ddots & a \\ & & & & b & a \end{bmatrix} \cdot \begin{Bmatrix} U_2 \\ U_3 \\ U_4 \\ \vdots \\ U_{Nc+1} \end{Bmatrix} = \mathbf{A} \cdot \begin{Bmatrix} U_2 \\ U_3 \\ U_4 \\ \vdots \\ U_{Nc+1} \end{Bmatrix} = 0 \quad (30)$$

with $a = -(Z_1 + Z_2 + Z_c)/G$ and $b = Z_1 + Z_2$. The eigenfrequencies of Eq. (30) can be calculated by letting the determinant of $\mathbf{A}$ equals to 0, which leads to

$$|\mathbf{A}| = a^{Nc} = 0 \quad (31)$$

Assuming the resistance $R_1$ and $R_2$ of shunting circuits to be zero, the natural frequency is obtained from Eq. (31) as

$$f_n = 1/(2\pi\sqrt{(L_1 + L_2) \cdot C_p}) = 4450 \text{ Hz} \quad (32)$$

It indicates that the natural frequency of the electrical transmission line maintains the same with that of the electrical cells. This is fundamentally different from two-way electrical transmission lines. As seen from Fig. 3(b), thanks to the resonance feature, the electrical system has the highest wave transmission capability around $f_n$.

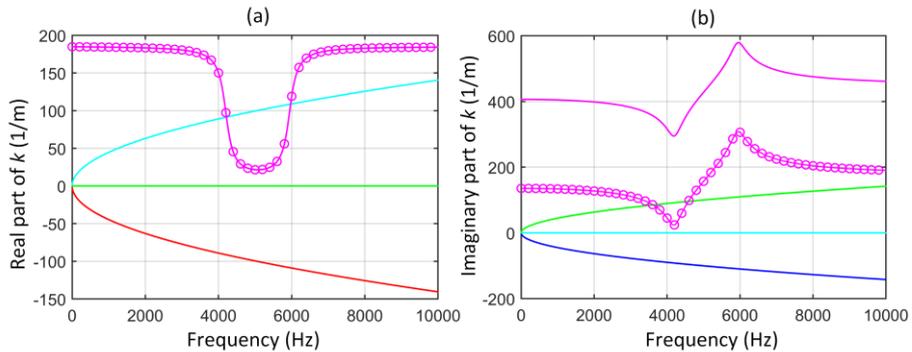

**Fig. 3.** Dispersion relations of the uncoupled electrical transmission line and mechanical beam: (a) real part of wavenumber $k$; (b) imaginary part of wavenumber $k$. Electrical wave modes: pink-solid line ($G$=-0.001), pink-solid line with circles ($G$=-0.1). Propagating elastic wave modes: cyan solid line and red solid line. Evanescent elastic wave modes: green solid line and blue solid line.

For the coupled electrical-mechanical cell, we firstly consider the scenario when $G$ is very small, $G$=-0.001 is chosen herein, the results of which are given in Fig. 4(a).

Due to the electrical-mechanical coupling effect, the dispersion curves are distinct from that of the uncoupled system, as presented in Fig. 3. Figure 4 (a.2) and (a.3) show that Im($k$) is almost the same for propagating elastic wave in opposite directions and appears to be large around the circuit natural frequency $f_n$, which indicates that local-resonance bandgaps exist in this system and the wave attenuation capability in the bandgaps are the same for opposite directions. These phenomena are due to the extremely small coupling intensity between neighboring electrical cells, which leads to little electrical energy transmission through the electrical system. Thus, the electrical circuits and the mechanical beam are interacted locally, viz. all electrical cells function as local resonators for the piezoelectric beam. It has been widely understood that reciprocal local-resonance bandgaps exist in such systems. Actually, as stated in Section 2.2, when $G$ equals to 0, the electrical-mechanical system would become a conventional local-resonance piezoelectric metastructure shunted with inductors ($L=L_1+L_2$) in series with resistors ($R=R_1+R_2$). Thus, the results of the case $G$=-0.001 corroborate with the known properties of local-resonance piezoelectric metastructures. It qualitatively verifies correctness of the obtained analytical dispersion equation Eq. (26).

When the one-way coupling coefficient $G$ is increased to $G$=-0.1, the dispersion curves of the electrical-mechanical cell are depicted in Fig. 4(b). It has been explained from Fig. 3 that the transmission capability of energy in the electrical transmission line is improved obviously in this scenario, thus the impact of which on the transmission of elastic wave cannot be ignored. From the dispersion curves presented in Fig. 4(b.2) and (b.3), it is found that, compared to the uncoupled scenario, obvious alteration of Im($k$) is observed around the natural frequency $f_n$ of the electrical transmission line, which indicates that the local-resonance effects are maintained for the two directions even with the existence of wave propagation in the electrical transmission line. Thanks to the non-zero Im($k$) around $f_n$, which implies wave attenuation, the system possess local-resonance bandgaps around $f_n$ in opposite directions. This is distinct from piezoelectric metamaterials with two-way electrical transmission lines shunting, where the bandgap property is characterized by dispersion veering and crossing effects [30,33,35]. As depicted in Fig. 4(b.2) and (b.3), due to the significant interaction between transmission of electrical energy and that of mechanical energy, especially around $f_n$, the shape alteration of Im($k$) curves of the propagating elastic wave and electrical wave is distinct from that shown in Fig. 4(a.2) and (a.3). The Im($k$) curves are not symmetric about the horizontal axis anymore, which implies appearance of different wave attenuation capability in opposite directions. Specifically, the attenuation capability in the backward direction is larger than that in the forward direction, therefore nonreciprocity exists in this scenario. Its mechanism is that the transmission of energy in the electrical transmission line is unidirectional, which thus manipulates the transmission of elastic energy direction-dependently. As can be seen in Fig. 3(b), around $f_n$, the electrical wave has the highest transmission capability. The ability to manipulate elastic wave by the electrical transmission line is thus dominated around $f_n$, which accordingly leads to strong nonreciprocity around $f_n$. Therefore, the achieved frequency range with nonreciprocity is narrowband. On the other hand, since it is based on the local-resonance characteristic, it has the advantage of subwavelength manipulation and could be implemented in the low-frequency range. From Eq. (32), we can also know that the nonreciprocity frequency can be tuned by simply changing the inductance values $L_1$ and $L_2$.

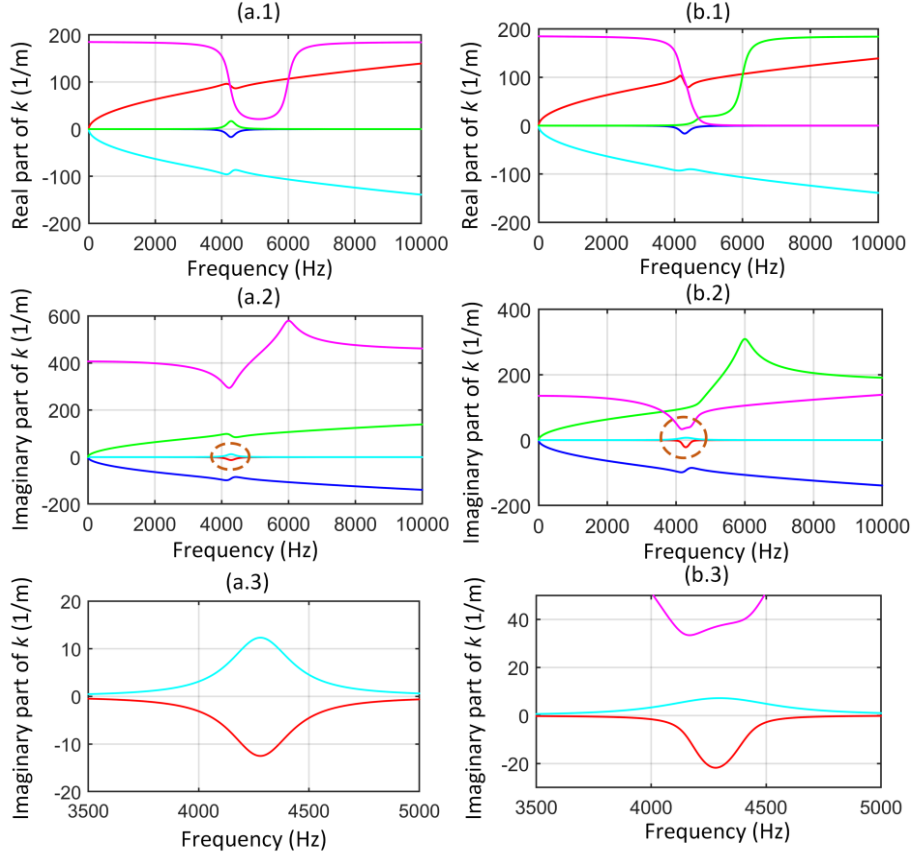

**Fig. 4.** Dispersion relations of the electrical-mechanical cell with different one-way electrical coupling coefficients $G$: (a) $G$=-0.001; (b) $G$=-0.1. Figures (a.1) and (b.1) present the real part of wave number $k$ while Figures. (a.2) and (b.2) present the corresponding imaginary parts of wavenumber $k$ with the same color lines. Figures (a.3) and (b.3) present the zoom-in views of the dashed-circle areas in Figs. (a.2) and (b.2), respectively. The dispersion curves of the uncoupled system shown as Fig. 3 can be employed as the comparing benchmark to elucidate the wave propagation behaviors .

## 4. Numerical simulations

In section 3, nonreciprocal behaviors of the presented electrical-mechanical system are qualitatively analyzed based on dispersion relations. To illustrate nonreciprocity of the system more intuitively and characterize its intensity quantitatively, we further utilize the FEM software Comsol Multiphysics to study wave transmission properties in the finite piezoelectric metastructure presented in Fig. 2(b). As stated in Section 2, the one-way electrical coupling element can be seen as a voltage-controlled voltage source (VCVS). Hence, in the Comsol FEM model, we employ VCVS in the *Electrical-circuit physics* module to represent the one-way coupling electrical element given in Fig. 1(b). The Rayleigh damping is applied to the mechanical structures and a weak damping ratio $\xi$ is assumed. The parameters of the piezoelectric metastructure are illustrated in Table 1 and Table 2.

**Table 2.** Parameters of the finite piezoelectric metastructure in numerical simulations

| Cell number $N_c$ | Beam length $d$ (mm) | PZT position $d_1$ (mm) | Actuating/sensing position $d_2$ (mm) | Damping ratio $\xi$ |
|---|---|---|---|---|
| 16 | 820 | 275 | 100 | $10^{-4}$ |

As shown in Fig. 2(b), the wave propagation behaviors in both directions are to be studied. The frequency response function (FRF) $a_s/F$ is defined to evaluate wave transmission behaviors, with $a_s$ being the acceleration amplitude of the measuring point and $F$ being the actuating force amplitude. For the presented metastructure with different circuit conditions, the calculated frequency response functions $a_s/F$ are depicted in Fig. 5. It is seen that when the electrical transmission line is disconnected with PZT patches, namely the "Open circuit" case, there exists no local-resonance bandgaps. This case is used as the comparing benchmark to indicate wave attenuation properties induced by the electrical system. When the electrical transmission line is shunted and its one-way coupling coefficient $G=0$, the system becomes a local-resonance piezoelectric metamaterial and thus the result of which is denoted as the "Local resonance" case in Fig. 5. It is observed that the local-resonance bandgap exits around the circuit natural frequency $f_n$, which coincides with the prediction by theories of local-resonance metamaterials. According to the analysis of Section 3, when the one-way coupling coefficient $G$ is increased, the forward and backward wave attenuation capability would be altered. Here we calculate frequency response functions $a_s/F$ under different nonzero $G$. The results in Fig. 5 indicate that, compared to the "Open circuit" case, wave attenuation exists around $f_n$ in the opposite directions for the cases with $G=$-0.05, -0.10 and -0.15, viz. the local-resonance bandgaps are maintained when the one-way electrical coupling feature is introduced. Additionally, it is seen that the backward attenuation capability is improved compared with the "Local-resonance" case while that in forward direction is weakened. Therefore, the wave attenuation capability of the opposite directions is distinguished and nonreciprocity emerges. Lastly, we find that if the one-way coupling coefficient $G$ is increased, the forward transmission in the bandgap is increased while the backward transmission is decreased. This feature entails that the asymmetry ratio of the presented nonreciprocal metastructure can be adjusted by changing the value of $G$.

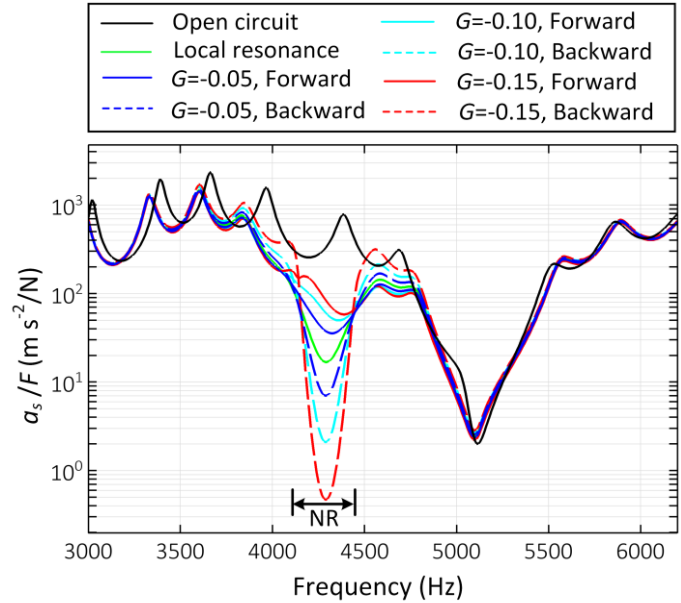

**Fig. 5.** Numerical frequency response functions (FRFs) $a_s/F$ of the piezoelectric metastructure: for the case "Open circuit", the electrical transmission line is disconnected with PZT patches; for the case "Local resonance", the one-way electrical coupling coefficient $G=0$ and the system becomes a conventional local-resonance piezoelectric metamaterial; for cases with different nonzero $G$, there exists one-way electrical coupling feature. Nonreciprocity (NR) frequency range is indicated in the figure.

To see how elastic wave decay along the piezoelectric metastructure when the one-way electrical coupling feature is employed, Figure 6 presents the distribution of transverse acceleration amplitude along the 16 metastructure cells, where the excitation frequency is chosen at $f_e$=4290 Hz within the bandgap. It is seen that, elastic wave decays more drastically in the backward direction for all cases, which implies larger backward wave attenuation capability. Moreover, if $G$ is increased, the backward transmitted elastic energy tends to be localized around the right end while the forward transmitted elastic energy can maintain high values in a larger distance. This verifies that, when the one-way electrical coupling coefficient $G$ is increased, the wave attenuation capability in backward direction is improved and that in the forward direction is reduced.

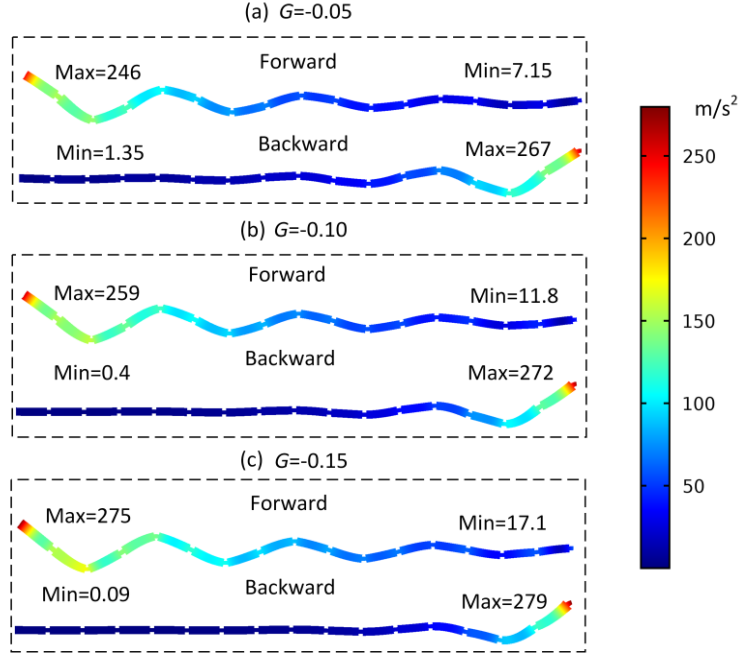

**Fig. 6.** Transverse acceleration amplitude distribution along the 16 cells of the piezoelectric metastructure, where the maximum and minimum values of acceleration amplitude are illustrated. Both the forward and backward wave transmissions are presented and the excitation frequency $f_e$ is designated at 4290 Hz within the bandgap.

We further calculate the time-domain acceleration responses of the piezoelectric metastructure. Three scenarios with different $G$ are considered: $G$=-0.05, $G$=-0.10, $G$=-0.15. The actuating and sensing conditions are the same with that in the frequency-domain analysis. The excitation frequency $f_e$=4290 Hz in the bandgap is chosen. The steady response in an interval of $\Delta t$ =0.01s is recorded and present in Fig. 7. It shows that the acceleration response of the measuring point is smaller when wave transmits backward. This demonstrates the existence of nonreciprocity in the proposed system. Also, the nonreciprocity is enhanced when the one-way electrical coupling coefficient $G$ is enlarged. These time-domain results corroborate with conclusions obtained in the frequency-domain analysis. In the time-domain analysis, it is also observed that the electrical-mechanical system becomes unstable if the one-way coupling coefficient $G$ excesses certain critical values. Actually, the instability issue exists commonly in nonreciprocal systems with active elements [18,20,21]. Thus, though increasing the coupling coefficient $G$ could lead to stronger nonreciprocity, we need to limit $G$ in the designing process so as to guarantee stability of the system.

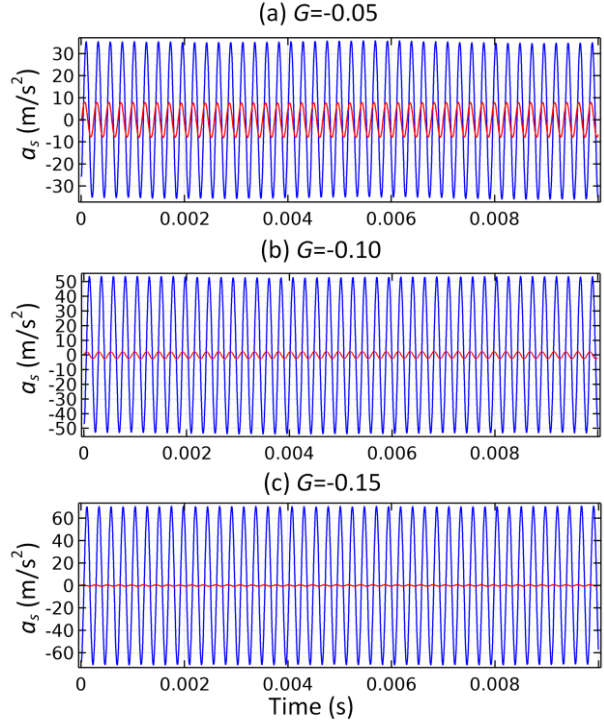

**Fig. 7.** Time-domain transverse acceleration responses of the sensing point indicating forward wave transmission (blue line, sensing point at position B) and backward wave transmission (red line, sensing point at position A). The excitation frequency $f_e$ is at 4290 Hz within the bandgap.

It is known that for local-resonance piezoelectric metamaterials, resistance of circuit resonators could impact the resonance intensity and hence the wave attenuation capability in the bandgap. It has been shown in Fig. 5 that the presented piezoelectric metastructure could maintain local-resonance bandgaps in both directions. To reveal how resistance of electrical cells influence the bandgap performance, the system with different values of $R_1$ and $R_2$ in the electrical transmission line are examined and the results are presented in Fig. 8. It implies that with the increasing of resistance of electrical cells, thanks to the weaker resonance intensity, the bandgap attenuation is reduced and accordingly transmission ratios in the bandgap frequency range are increased for both directions. In addition, though nonreciprocity is preserved when larger resistance is employed, its asymmetric transmission ratio is decreased. These results verify the adaptiveness of nonreciprocity with the resistance of electrical cells.

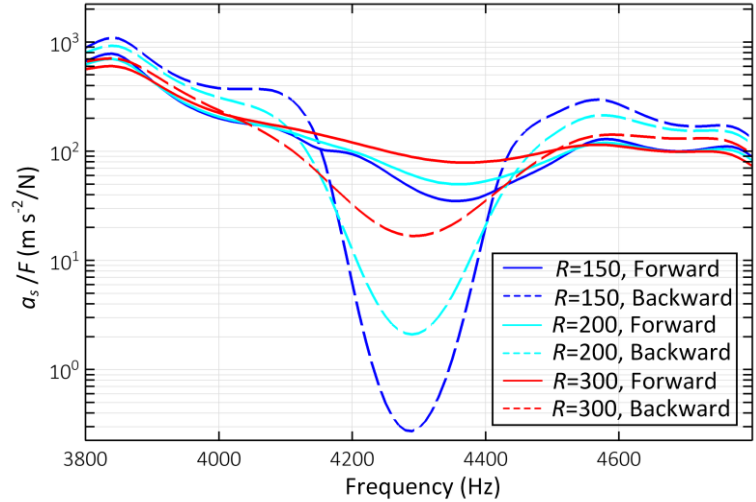

**Fig. 8.** Numerical frequency response functions (FRFs) $a_s/F$ of the piezoelectric metastructure with different resistance in electrical cells, $R_1=R_2=R$.

In engineering applications, the tunability of nonreciprocity is a key factor worth consideration. According to the results shown in Fig. 5, the asymmetry ratio of this nonreciprocal system is tunable with the one-way coupling coefficient $G$, which can be easily changed by altering the resistance $R_a$ and $R_b$ of the inverting amplifier as seen in Fig. 1(b). In addition, based on the results shown in Fig. 8, we can also change the resistance $R_1$ and $R_2$ to tune the nonreciprocal wave attenuation. Since the achieved nonreciprocity is closely associated with the local-resonance frequency $f_n$ of the electrical transmission line, we are able to tune the nonreciprocity frequency by adjusting the inductance $L_1$ and $L_2$. The synthetic inductors such as the Antonio's circuit [36] can be used to implement these inductors physically, which are easily tunable. The circuit design shown in Fig. 2 is based on physical electrical elements. It is also a feasible scheme to employ digital circuits [24,37] to implement this design, which could render the achieved nonreciprocity programmable.

The practical metastructures all consists of finite number of cells. Therefore, even within the bandgaps, a certain amount of energy can still propagate through the system, which depends on the wave attenuation capability. The presented piezoelectric metastructure can break the asymmetry of wave attenuation in opposite directions and thus achieves nonreciprocity. Its benefit is that, by simply varying the wave attenuation ratios, the transmitted wave energy density of opposite directions can be tuned nonreciprocally with adaptive control of the asymmetry ratio. The simulation results presented in Fig. 7 reveals that the asymmetry ratio, which is defined as the ratio of forward FRF to backward FRF, can be tuned to be relatively small as 4.5, and also as large as 141 at the frequency 4290 Hz. In this regard, the presented system provides us with more flexibility to manipulate the wave propagation behaviors than other nonreciprocal systems such as spatial-temporal metamaterials [12,13], of which the bandgap frequency ranges in opposite directions are distinct and wave transmission is either prohibited or allowed in the nonreciprocity frequency range.

## 5. Experimental investigations

To verify the observed nonreciprocity in the analytical and numerical investigations, an experimental set-up of the piezoelectric metastructure is built and wave transmission behaviors in the opposite directions are tested. The testing scheme is presented as Fig. 9(a) and the experimental photograph of the piezoelectric metastructure is presented as Fig. 9(b). The one-way electrical transmission line is configured in breadboards. For the convenience, instead of applying point loads and measuring acceleration responses as discussed in Section 4, we employ one PZT pair to actuate the beam near one end and another PZT pair to measure the output response near the other end. The actuating voltage produced by a signal generator is amplified by the voltage amplifier and then input to the actuating PZT pair. As shown in Fig. 9(a), the actuating voltage $U_i$ is employed to characterize wave intensity input to the metastructure. The sensing PZT pair is shunted with a resistor $R_s$, and the voltage response $U_o$ across which is employed to characterize the wave intensity transmitted through the metastructure. Then the frequency response functions (FRFs) defined as $U_o/U_i$ are measured to reflect the wave transmission capability. Figure. 9(a) only presents the schematic of measuring forward wave transmission; when the backward wave transmission is to be measured, the PZT pair at position A is used for sensing while the PZT pair at position B is used for actuating.

The physical and geometrical parameters of the piezoelectric metastructure are presented as Table 1 and Table 2. The parameters of electrical elements in the experiment are measured as follows. The inductance $L_1$ and $L_2$ are around 0.095 H, and their inherent DC resistance $R_1$ and $R_2$ are around 167 Ω. The capacitance of PZT patches in each unit cell is $C_p$=6.5 nF. The experimental natural frequency of the electrical transmission line can then be calculated as $f_n^e$=4529 Hz, which is a little different from the theoretical one $f_n$=4450 Hz as presented in Eq. (32). The resistance in the one-way coupling electrical element presented as Fig. 1(b) is $R_a$=1 MΩ while the resistance $R_b$ is chosen as: $R_b$=0 for the case $G$=0; $R_b$=100 kΩ for the case $G$=-0.1; $R_b$=200 kΩ for the case $G$=-0.2. The resistance connected with the measuring PZT pair is $R_s$=10 kΩ.

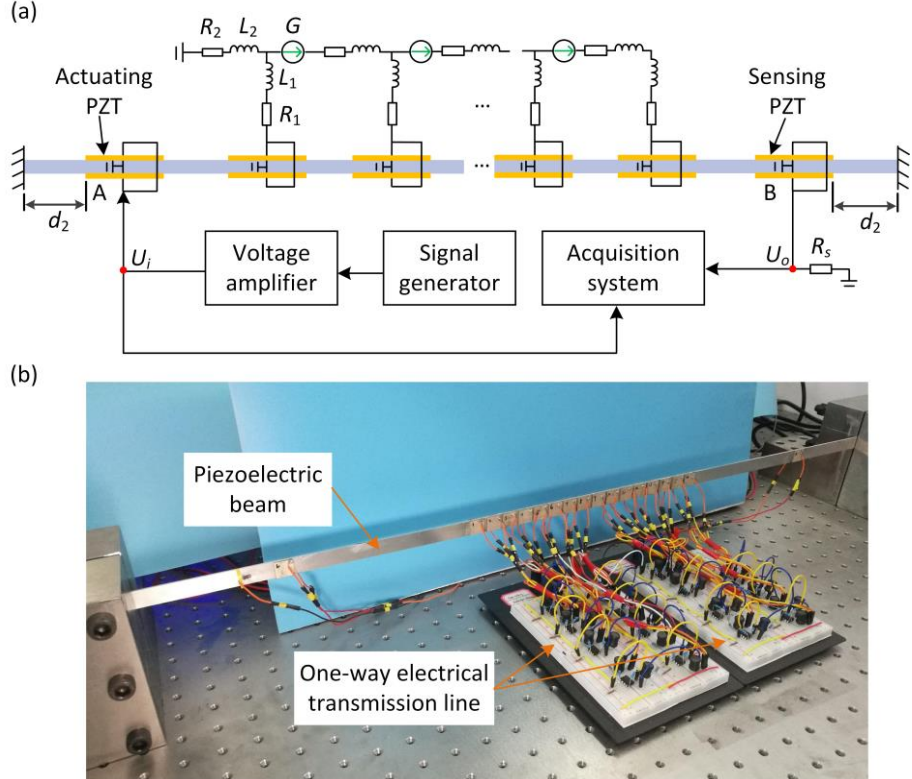

**Fig. 9.** (a) Scheme of measuring forward elastic wave transmission in the piezoelectric metastructure (for backward transmission, PZT at position A is used for sensing and PZT at position B is used for actuating). (b) Photograph of the experimental set-up

The frequency response functions (FRFs) $U_o/U_i$ are firstly measured for the two opposite directions. Four scenarios with different circuit conditions are considered, the results of which are shown in Fig. 10: "Open circuit" case, where the electrical transmission is disconnected with PZT patches; "Local resonance" case, where $G$=0; $G$=-0.1; $G$=-0.2. The sweeping harmonic voltage signal is employed to actuate the beam. The obtained results show that for the "Open circuit" and the "Local resonance" cases, the wave transmission curves for the two opposite directions are in well coincidence. It verifies that the experimental piezoelectric metastructure has ideal symmetry if the one-way coupling feature is not introduced, which therefore can ensure that if there exists any transmission difference for opposite directions, it is not due to the structure asymmetry. When $G$=0, the local-resonance bandgap is observed around the natural frequency of electrical cells $f_n^e$=4529 Hz, as could be theoretically predicted. When the one-way coupling coefficient $G$ is increased to $G$=-0.1 and $G$=-0.2, the bandgaps still exist while transmission ratios in the bandgap are distinct for opposite directions. The wave attenuation in backward direction is larger than that in forward direction, which indicates the existence of nonreciprocity. Also, the nonreciprocity phenomenon is enhanced when $G$ is increased from $G$=-0.1 to $G$=-0.2. These experimental results qualitatively agree with numerical results in Section 4, and can demonstrate the existence of nonreciprocity phenomenon in the proposed electrical-mechanical system.

The time-domain responses of the piezoelectric metastructure are also measured. A harmonic voltage with the constant amplitude is employed to actuate the beam. The excitation frequency $f_e$ is fixed at $f_e$=4650 Hz within the bandgap. The voltage responses $U_o$ of the sensing PZT are recorded and presented in Fig. 11, where the four testing cases have identical circuit conditions with that in Fig. 10. In each subfigure, the voltage responses $U_o$ indicating forward (blue line) and backward (red line) wave transmission capability are shown. It implies that, for the "Open circuit" and "Local resonance" cases, the amplitude of $U_o$ in opposite directions are almost the same. And the voltage amplitude of the "Local-resonance" case is reduced compared to that of the "Open circuit" case, which is resulted from the local-resonance bandgap effects. For the case "$G$=-0.1", the forward and backward transmissions appear to be distinct, thus indicating existence of nonreciprocity. When a greater $G$ value is used, $G$=-0.2 herein, the nonreciprocity is also improved. In this case, the backward transmission is reduced by 52 dB compared to the forward transmission, which means that the nonreciprocity intensity of the system is very high. The conclusions achieved from time-domain results agree with that from the frequency-domain results as presented in Fig. 10 and can thus verify the correctness with each other.

In the experiment, it is found that if the one-way coupling coefficient $G$ is too large, the electrical-mechanical system becomes unstable, which is also elucidated in numerical simulations. Therefore, though a larger value of $G$ could lead to higher nonreciprocity intensity, we need to carefully design $G$ to avoid instability of the system. For increasing the maximum allowable value of $G$, we can add damping layers to the host structure to improve the stability of the system. In the reference [38], it is proposed to employ electrical filters to stabilize the active metamaterial. This technique may also be utilized to guarantee stability of the presented piezoelectric metastructure and therefore enhance nonreciprocity intensity.

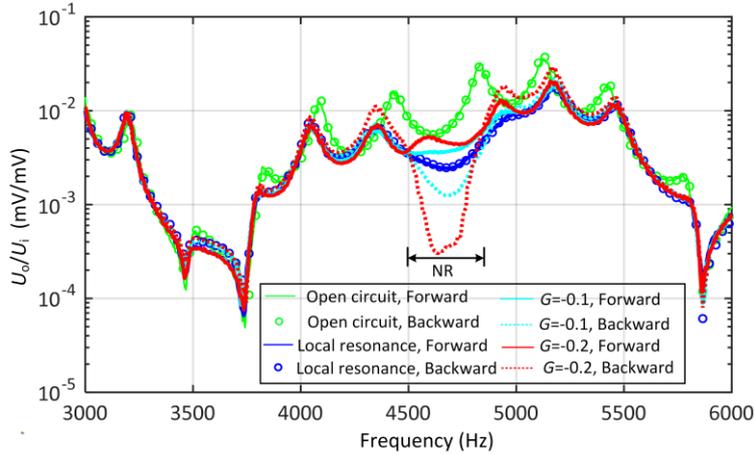

**Fig. 10.** Experimental frequency response functions $U_o/U_i$ in opposite directions of the piezoelectric metastructure: for the case "Open circuit", the electrical transmission line is disconnected with PZT patches; for the case "Local resonance", the one-way electrical coupling coefficient $G$ equals to 0; for cases with $G$=-0.1 and -0.2, the one-way electrical coupling feature exits. The nonreciprocity (NR) frequency range is indicated in the figure.

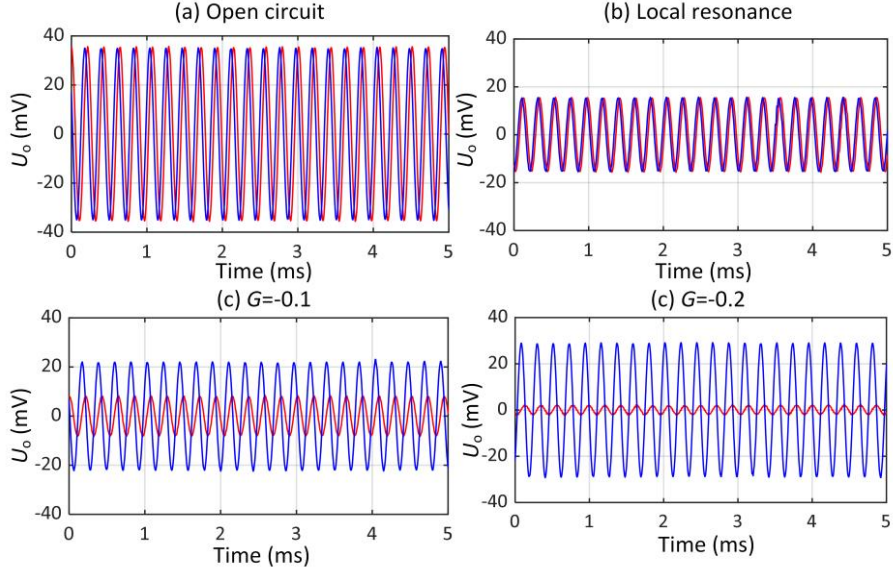

**Fig. 11.** Experimental time-domain voltage responses $U_o$ indicating forward wave transmission (blue line, sensing PZT at position B) and backward wave transmission (red line, sensing PZT at position A). The excitation frequency is $f_e$=4650 Hz within the bandgap.

## 6. Conclusions

In this paper, we report nonreciprocal wave transmission properties in a piezoelectric metastructure with a one-way electrical transmission line shunting. We theoretically and experimentally demonstrate the existence of nonreciprocity in the presented electrical-mechanical system. It is found that the piezoelectric metastructure is able to maintain local-resonance bandgaps in opposite directions but with distinguished wave attenuation capability, which therefore leads to nonreciprocity of elastic wave. Specifically, compared to the conventional piezoelectric metamaterials with same parameters, the wave attenuation capability in one direction of the proposed system is reduced while it is improved in the other direction. This nonreciprocal wave attenuation phenomenon is fundamentally distinct from wave propagation behaviors in piezoelectric metamaterials shunted with two-way electrical transmission lines. We also find that the asymmetry ratio of the presented nonreciprocal system is adaptive with the one-way electrical coupling coefficient as well as the resistance of electrical cells. Overall, the proposed system creates a new means to realize linear and adaptive nonreciprocity with easy implementations.

The nonreciprocity mechanism in this paper is based on local-resonance bandgaps. Therefore, the achieved frequency range with nonreciprocity is narrowband. We can utilize the methods that that have been proposed to increase width of local-resonance bandgaps, such as the rainbow effect [39] and the negative-capacitance circuit [40], to achieve broadband nonreciprocity. On the other hand, the presented nonreciprocity mechanism might also be attainable in Bragg-type bandgaps, which is broadband, but needs further explorations in the future. In addition to the one-dimensional system in this research, the one-way electrical coupling feature could also be introduced to manipulate elastic wave transmission in 2D electrical-mechanical systems. Lastly, we envision that the idea of employing nonreciprocity of electrical wave to induce nonreciprocity of elastic wave could possibly be extended to explore interactions of other nonreciprocal physical media with reciprocal elastic media.

**CRediT authorship contribution statement**

**Yisheng Zheng**: Conceptualization, Methodology, Software, Formal analysis, Investigation, Writing-Original Draf. **Junxian Zhang**: Investigation, Writing-Original Draft. **Yegao Qu**: Resources, Writing - Review & Editing, Supervision. **Guang Meng**: Resources, Writing - Review & Editing, Supervision.

**Acknowledgements**

This work was supported by the National Natural Science Foundation of China (No. 12002203; Grant No. 11922208; 11932011); and the China Postdoctoral Science Foundation (No. 2020M671105); Natural Science Foundation of Shanghai, China (Grant No. 18ZR1421200).